\documentclass[preprint,showpacs,preprintnumbers,amsmath,amssymb,superscriptaddress]{revtex4}

\usepackage{amssymb}
\usepackage{txfonts}

\usepackage{graphicx}
\usepackage{dcolumn}
\usepackage{bm}
\usepackage{subfigure}

\begin{document}


\title{Voltage-controlled intracavity electromagnetically induced transparency\\ with asymmetry quantum dots molecule}

\author{Yandong Peng}
\affiliation{State Key Laboratory of High Field Laser Physics,
Shanghai Institute of Optics and Fine Mechanics, Chinese Academy of
Sciences, Shanghai 201800, China}%
\affiliation{Graduate University of Chinese Academy of Sciences,
Beijing 100049, China}
\author{Yueping Niu}\thanks{Corresponding author.
E-mail: niuyp@mail.siom.ac.cn}%
\affiliation{State Key Laboratory of High Field Laser Physics,
Shanghai Institute of Optics and Fine Mechanics, Chinese Academy of
Sciences, Shanghai 201800, China}%
\author{Ni Cui}
\affiliation{State Key Laboratory of High Field Laser Physics,
Shanghai Institute of Optics and Fine Mechanics, Chinese Academy of
Sciences, Shanghai 201800, China}%
\affiliation{Graduate University of Chinese Academy of Sciences,
Beijing 100049, China}
\author{Shangqing Gong}\thanks{Corresponding author.
E-mail: sqgong@mail.siom.ac.cn; \ Tel: 086-021-69918163; \ Fax: 086-021-69918021}%
\affiliation{State Key Laboratory of High Field Laser Physics,
Shanghai Institute of Optics and Fine Mechanics, Chinese Academy of
Sciences, Shanghai 201800, China}%

\date{\today}

\begin{abstract}
We theoretically investigate the phenomenon of voltage-controlled
intracavity electromagnetically induced transparency with asymmetric
double quantum dot system. The impact of voltage on frequency
pulling and cavity linewidth narrowing are discussed. The linewidth
and position of the cavity transmission can be engineered by the
bias voltage. The scheme may be useful in integrated electro-optical
device in quantum information process.
\end{abstract}

\maketitle

\noindent Electromagnetically induced transparency (EIT) \cite{EIT,
RevMod}, as a useful technology, has received much attention for its
potential applications in a wide domain, such as enhancement of
nonlinear optical processes \cite{harris}, quantum coherent control
\cite{lukinRevMod}, and quantum information and memory
\cite{lukinQumIn,Fleis,Duan}. An EIT medium placed in a cavity can
substantially affect the properties of the resonator system and
significantly narrow the cavity linewidth, which is known as
intracavity-EIT termed by Lukin and co-workers \cite{lukin ol}.
Later, many EIT-based works have been carried out in optical
resonator with atomic system, such as optical bistability and
multistablity \cite{bistability}, photon-photon interaction
\cite{imamoglu}, slow light \cite{zhu1, Hau}, cavity-linewidth
narrowing by EIT with hot \cite{xiao1} and cold \cite{zhu2}
atomic-Rb system respectively, and so on.

Quantum dot (QD) is three-dimensional confined semiconductor
nanostructure in which electrons and holes exhibit discrete energy
levels \cite{gammon}. This atomic-like property allows us to treat
dynamical evolution of a QD system with similar methods in atomic
physics. Unlike atoms, they can be customized as expected. Moreover,
a QDs system has many advantages than an atomic system, such as high
nonlinear optical coefficients, large electric-dipole moments and
ease of integration, which make it as a promising candidate for
quantum information processing. Great progresses in the fabrication
and physics of single quantum dots focus people's attention on
coupled quantum dots \cite{prl06}. For example, an asymmetric
quantum dot molecule (QDM) consists of two QDs with different band
structures which is coupled by tunneling. A near resonance incident
light can excite an electron from the valence to the conduction band
in a dot, and the electron can tunnel to the other dot. The
tunneling effect is sensitive to external bias voltage, then the
interdot oscillations can be controlled by the applied voltage. Many
works have been carried out about the asymmetric QDM system, such as
coherent control of electron tunneling \cite{prb04}, optical
signature of asymmetric QDM \cite{stinaff04}, voltage-controlled
slow light \cite{yuan} and pulse storage and retrieval \cite{li},
etc. The flexible QDM combined with the excellent optical and
electrical technology exhibits an attractive prospect in integrated
electro-optical devices in quantum information process.

Optical cavity is a useful tool for investigation of light-matter
interaction and provides us many principle demonstrations which
inspires us to extend these results to other fields. In recent
years, much attention has been paid to the investigation of a
quantum dot-cavity system. These include cavity-coupled resonance
fluorescence \cite{muller}, strong-coupling interaction
\cite{imamoglu2, pho-correlation, rabi-splitting}, dark-polarization
soliton \cite{soliton}, quantum bistability \cite{qu-bistability},
cavity-assistance generation of entangled photon \cite{03prb,
09prb1, 09prb2}, and so on. Some of us have also studied
transmission spectrum of a double quantum-dot-nanocavity system in
photonic crystals \cite{qian}.

Intrigued by these studies, in this paper we combine asymmetric QDM
sample with optical resonator and theoretically describe the
phenomenon of an intracavity-induced transparency controlled by
voltage. When a QDM sample is placed inside an optical resonator,
the cavity field can couple its interband transition. In the
presence of the bias voltage, the interdot tunneling is enhanced,
which affects its optical property. Then voltage-controlled
tunneling replace pump laser coupling the interdot transition, and
the two transitions consist of an EIT configuration \cite{prl06}.
Therefore, we can control the intracavity EIT by voltage, which
results in frequency pulling and cavity-linewidth narrowing. The
asymmetric QDM can be simplified as the model in Fig. 1(a). The
ground state $|0\rangle$ is the system without excitation and the
exciton state $|1\rangle$ has a pair electron and hole bound in the
first dot. The indirect exciton state $|2\rangle$ contains one hole
in the first dot and an electron in the second dot, and it is
coupled with $|1\rangle$ by the electron tunneling. A probe field
couples $|0\rangle \leftrightarrow |1\rangle$ transition with Rabi
frequency $g= - \mu _{10} E_p /2\hbar $, where the electric dipole
moment $\mu _{10}$ denotes the coupling to the radiation field of
the excitonic transition, and $E_p$ is the amplitude of the probe
field.

According to the standard procedure \cite{scully}, we get the
dynamical equations of the QDM sample in the interaction picture and
in the rotating wave approximation:
\begin{eqnarray}
\dot \rho _{00}  = \gamma _{10} \rho _{11}  + \gamma _{20} \rho
_{22}  + ig(\rho _{01}  - \rho _{10} ),\nonumber\\
 \dot \rho _{11}  =  - \gamma
_{10} \rho _{11}  + iT_e(\rho _{12}  - \rho _{21} ) + ig(\rho _{10}
- \rho _{01} ),\nonumber\\
 \dot \rho _{22}  =  - \gamma _{20} \rho
_{22}  +
iT_e(\rho _{21}  - \rho _{12} ) ,\nonumber \\
 \dot \rho _{10}  =  - (i\Delta  +
\Gamma _{10} )\rho _{10}  - ig(\rho _{00}  - \rho _{11} ) - iT_e\rho
_{10}, \\
 \dot \rho _{20}  =  - [i(\Delta  - \omega _{12} )
+ \Gamma
_{20} ]\rho _{20}  + ig\rho _{21}  - iT_e\rho _{10},\nonumber \\
\dot \rho _{12} =  - (i\omega _{12}  + \Gamma _{12} )\rho _{12}  -
iT_e(\rho _{22}  - \rho _{11} ) - ig\rho _{02},\ \ \ \ \ \ \nonumber
\end{eqnarray}
together with $\rho _{kj}^ *=\rho _{jk}$ and the closure relation $
\sum\nolimits_j {\rho _{jj} }  = 1 \left( {j,k \in  \left\{ {0,1,2}
\right\}} \right)$. Here ${\gamma _{jk} }$ is lifetime broadening
linewidth, and ${\Gamma _{jk} }$  is dephasing broadening linewidth
which is usually the dominant mechanism and can be controlled by
adjusting barrier thickness in QDM. The detuning
$\Delta=\omega_{01}-\omega_p$ and the energy difference
$\omega_{12}=\omega_{02}-\omega_{01}$. $T_e$ is the
electron-tunneling matrix element. In the following, all parameters
are scaled by ${\Gamma _{10} }$, which is about in the order of meV
for (In,Ga)As/GaAs \cite{prl06, stinaff04}.

Consider the QDM system being initially in the ground state
$|0\rangle$, $\rho _{00}^{(0)}  = 1$ and $ \rho _{11}^{(0)} = \rho
_{22}^{(0)} = \rho _{21}^{(0)}  = 0$, and weak field approximation.
By solving the Eqs. (1), the linear susceptibility is given
by\cite{yuan}:
\begin{eqnarray}
\chi \left( \omega  \right) = \chi ' + i\chi '',\\
\chi '{\text{ = }}\left\{ { - \Delta \Gamma _{{\text{20}}}  -
(\Delta  - \omega _{12} )[T_e^2  - \Delta (\Delta  - \omega _{12}
)]} \right\}/\mathcal {D}
,\nonumber\\
\chi '' = \left\{ { - \Gamma _{{\text{10}}} (\Delta  - \omega _{12}
)^2  - \Gamma _{{\text{20}}} (\Gamma _{{\text{10}}} \Gamma
_{{\text{20}}}  + T_e^2 )} \right\}/\mathcal {D}
,\nonumber\\
\mathcal {D} = (\Gamma _{{\text{10}}} \Gamma _{{\text{20}}}  -
\Delta ^2  + \Delta \omega _{12}  + T_e^2 )^2  + [\Delta \Gamma
_{{\text{20}}}  + (\Delta  - \omega _{12} )\Gamma _{{\text{10}}}
]^2, \nonumber
\end{eqnarray}
where $T_e$ and $\omega_{12}$ can be simply tuned with bias voltage
\cite{prb04}. It is easy to see that, when the probe field detuning
is zero $(\Delta=0)$ and the bia voltage turns off $(T_e=0)$, the
real part of the linear susceptibility $\chi'$ is zero but its image
part still exists $(\chi''\neq 0)$. It indicates there is a big
absorption around the resonance frequency $\omega_{01}$, so the
probe field is absorbed. When the tunneling $T_e$ increases, EIT
arises and a transparency window appears at the position where
$\chi''$ is zero \cite{yuan}, which requires $\Delta \approx
\omega_{12}$ according to the material parameters in Ref.
\cite{kim}. It means that the position of the transparency window
varys with $\omega_{12}$. This provides the feasibility of
engineering the QDM response at different frequencies simply by
modulating the bias voltage.

Here, we consider an optical ring cavity of length $L$ with an
asymmetric QDM sample of length $l$. The real part $\chi'$ of the
susceptibility of the medium brings dispersion and additional phase
shift, and the imaginary  part $\chi''$ introduces absorption
leading to the attenuation of the field's amplification. When an
empty cavity resonance frequency $\omega_{c}$ is near to the EIT
frequency $\omega _{01}$ of the QDM, the resonance frequency $\omega
_r$ of the cavity+QDM system is governed by \cite{lukin ol}
\begin{eqnarray}
\omega _r  = \frac{1} {{1 + \xi }}\omega _c  + \frac{\xi } {{1 + \xi
}}\omega _{01},
\end{eqnarray}
where $\xi  = \omega _{01} ({l \mathord{\left/
 {\vphantom {l {2L}}} \right.
 \kern-\nulldelimiterspace} {2L}})(\partial \chi '/\partial \omega _p) $
 describes dispersion
 changes as a function of probe frequency. By considering
$\Gamma_{20} \ll \Gamma_{10} $ \cite{kim}, the dispersion in the
transparency window $(\Delta=\omega_{12})$ is approximately given by
\begin{eqnarray}
\frac{{\partial \chi '}} {{\partial \omega _p }} \approx \frac{{ -
T_e^4 + T_e^2 (\Gamma _{20}  - 2\Gamma _{10} \Gamma _{20}  + 2\Gamma
_{20} \omega _{12}^2 )}} {{(T_e^3  + 2T_e \Gamma _{10} \Gamma _{20}
)^2 }}.
\end{eqnarray}
To a given system, the dispersion strongly depends on the tunneling.
Without the QDM sample, the resonance frequency is the empty-cavity
frequency. When the QDM sample is placed in the resonator and the
tunneling turns on, $\omega_r$ is pulled strongly to $\omega_{01}$.
By changing the bias voltage, the tunneling could be enhanced or
suppressed, which leads to different degree of frequency pulling.
This pulling effect can be represented quantitatively by the
stabilization coefficient $\xi$. To get a better
frequency-stabilization effect, we just need to tune the bias
voltage to a proper value.

It is instructive to examine the modified cavity linewidth
\cite{lukin ol}
\begin{eqnarray}
\Delta \omega  = \frac{{\Delta \omega _0 (1 - r\kappa )}} {{\sqrt
\kappa  (1 - r)}}\frac{1} {{1 + \xi }},
\end{eqnarray}
where $\Delta \omega _0$ is an empty-cavity linewidth and  $ \kappa
= \exp \left( { - k l\chi''} \right)$ is the medium absorption per
round trip. Without tunneling, the QDM sample becomes a two-level
system. Large dispersion ${{\partial \chi' } \mathord{\left/
 {\vphantom {{\partial \operatorname{Re} (\chi )} {\partial \omega }}} \right.
 \kern-\nulldelimiterspace} {\partial \omega }}$
 goes with big absorption, and the probe field is absorbed. However, if the tunneling turns on,
 EIT occurs with larger dispersion and weak (or zero) absorption
 which much narrows the cavity linewidth. At the transparency window,
 Eq. (5) reduces to $\Delta \omega  = \Delta \omega _0 /(1 + \xi )
$. Since the coefficient $\xi$ depends on the tunneling, then we can
control the cavity linewidth by the bias voltage. Under the
different strength of the tunneling, we examine the modified cavity
transmission. The rate of the cavity linewidth, $\Delta \omega$ and
$\Delta \omega '$,
 under the different tunneling reads
\begin{eqnarray}
\frac{{\Delta \omega }} {{\Delta \omega '}} = \frac{{1 + \omega '_r
({l \mathord{\left/
 {\vphantom {l {2L}}} \right.
 \kern-\nulldelimiterspace} {2L}})[(\partial \chi ')'/\partial \omega _p ]}}
{{1 + \omega _r ({l \mathord{\left/
 {\vphantom {l {2L}}} \right.
 \kern-\nulldelimiterspace} {2L}})(\partial \chi '/\partial \omega_p )}} \approx \frac{{(\partial \chi ')'/\partial \omega _p }}
{{\partial \chi '/\partial \omega_p }}.
\end{eqnarray}
It is easy to see that the cavity linewidth is reversely
proportional to the dispersion. Different tunneling controlled by
bias voltage leads to different dispersion in transparency window,
then different linewidths of cavity transmissions can be achieved.

Figure 2 shows the cavity transmission under different conditions
and all parameters are scaled by $\Gamma_{10}$ for simplicity. Empty
cavity transmission is plotted in Fig. 2(a). When the asymmetric QDM
sample is embedded in the cavity but the bias voltage turns off, the
QDM sample can be considered as two-level system, then the probe
field is absorbed [see Fig. 2(b)]. Whereas, if the bias voltage
turns on, the tunneling increases and the QDM sample becomes an EIT
system. So we get a narrow transmission peak [see Fig. 2(c)]. Here,
the bias voltage replaces the pump field coupling the interdot
transition and EIT occurs in the asymmetric QDM sample. Larger
dispersion appears in the transparency window and it much narrow the
cavity transmission. Since the cavity linewidth is inverse to the
dispersion around the transparency window, when the dispersion
decreases by tuning the bias voltage, the cavity transmission
becomes broad [see Fig. 2(e)]. Then we achieve the
voltage-controlled intracavity EIT. Moreover, the narrow
transmission peak shifts with the variation of $\omega_{12}$ [see
Fig. 2(d)], which can also be simply controlled by the bias voltage.
That is to say that we can engineer the cavity response by the QMD
sample at a broad frequency range.

Also, we present a 3D demonstration of the variation of the
dispersion with $T_e$ and $\Gamma_{10}$ in Fig. 3. It shows that the
dispersion changes with $T_e$ in different QDM vary widely. When the
decoherence ($\Gamma_{10}$) is small, fine-tuning voltage in the
magnitude range 0 $\thicksim$ 0.5 leads to the dramatic change of
the dispersion. While, to a QDM sample with large decoherence, this
change tends to gentle . Then, as required, we can prepare an
appropriate QDM sample or control its decoherence by tuning its
temperature.

Experimentlly the self-assembled lateral QDM can be grown by
molecule beam epitaxy combing with \emph{in situ} atomic layer
precise etching \cite{03_05prb}. In the following calculation, the
material parameters consult Ref. \cite{yuan, kim}. The surface
density is $4\times10^{10} cm^{-2}$ with the optical confinement
factor $\Gamma=6\times10^{-3}$. The laser wavelength is 1.36$\mu$m
and $|\mu_{10}|/e=21 Å.$ The linewidth value for $\hbar\Gamma_{10}$
are 6.6 $\mu$ev and $\Gamma_{20}=10^{-4}\Gamma_{10}$. At $T_e=0.01$,
the calculated dispersion from Eq. (5) is -4.5$\times 10^{3}$, which
means we could get the cavity-linewidth narrowing factor of $10^3$.
This may be useful in weak-electricity-controlled high-resolution
spectroscopy and laser frequency stabilization. What's more, when
the bias voltage turns on / off, the probe field transmits / is
absorbed. This may be used to measure small electric field and also
as an electro-optic switch.

In conclusion, we theoretically discussed the effect of
voltage-controlled EIT to a resonator with the asymmetric QDM
sample. Frequency pulling and cavity-linewidth narrowing are
discussed. The voltage-controlled tunneling replaces pump laser
coupling interdot transition, which composes an EIT process with the
probe transition. An appropriate tunneling leads to larger
dispersion which much narrows cavity linewidth. Moreover, by tuning
the external voltage, we can engineer cavity response at different
frequency. This scheme combines excellence optical and electrical
technology, and the voltage-controlled EIT may have potential
application in integrated electro-optical device in quantum
information process.

This work was supported by the National Natural Science Foundation
of China (Grant Nos. 10874194, 60708008, 60978013, 60921004).

\clearpage
\section*{List of Figure Captions}
\noindent Fig. 1. Schematic band structure and level configuration
of an
   asymmetric quantum dot molecule (QDM) system for electromagnetically induced
   transparency process. An input laser field excites an electron from
   the valence band to the conduction band in a dot that can tunnel to the other dot.\\

\noindent Fig. 2. Cavity transmission as a function of probe-field
detuning $\Delta$.
   (a) Empty-cavity transmission. Cavity+QDM transmission for (b) $T_e=0$, $\omega_{12}=0$;
   (c) $T_e$=0.5, $\omega_{12}=0$; (d) $T_e$=0.5, $\omega_{12}=0.2$; (e) $T_e$=1, $\omega_{12}=0$.\\

\noindent Fig. 3. Dispersion at EIT window as a function of
tunneling $Te$
   in different QDM samples whose linewidth vary from $6\mu$ev to $50\mu$ev
   assuming $\Gamma_{20}=10^{ - 4}\Gamma_{10}$. For a better view, the dispersion
   above 7 are not been plotted.


\clearpage
Figure 1
\begin{figure}[htbp]
  \centering
  \includegraphics[width=9cm]{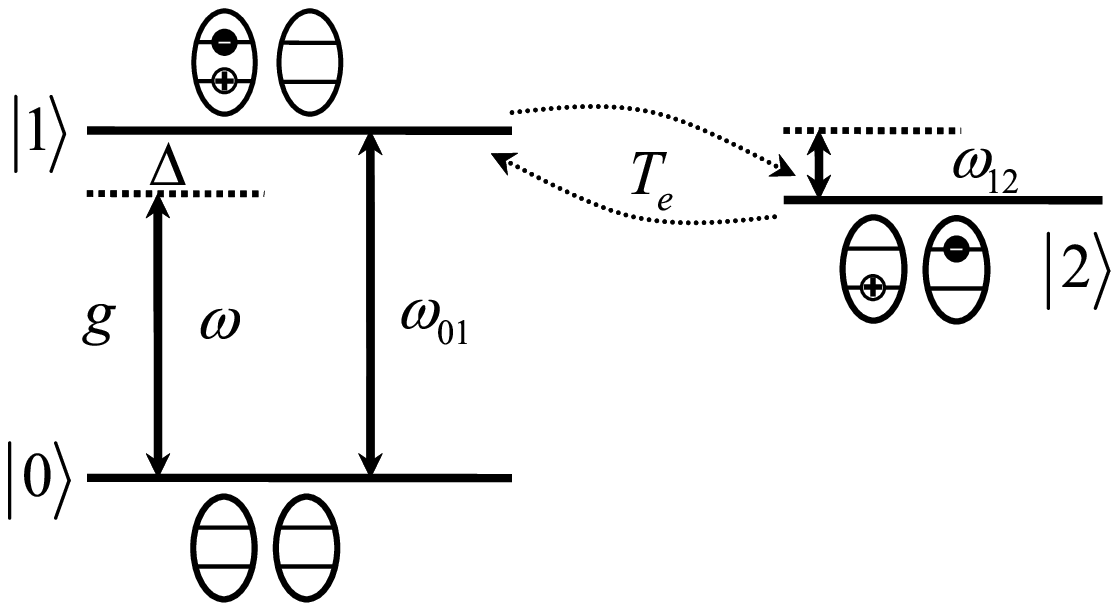}
\end{figure}

\clearpage Figure 2
\begin{figure}[htbp]
  \centering
  \includegraphics[width=9cm]{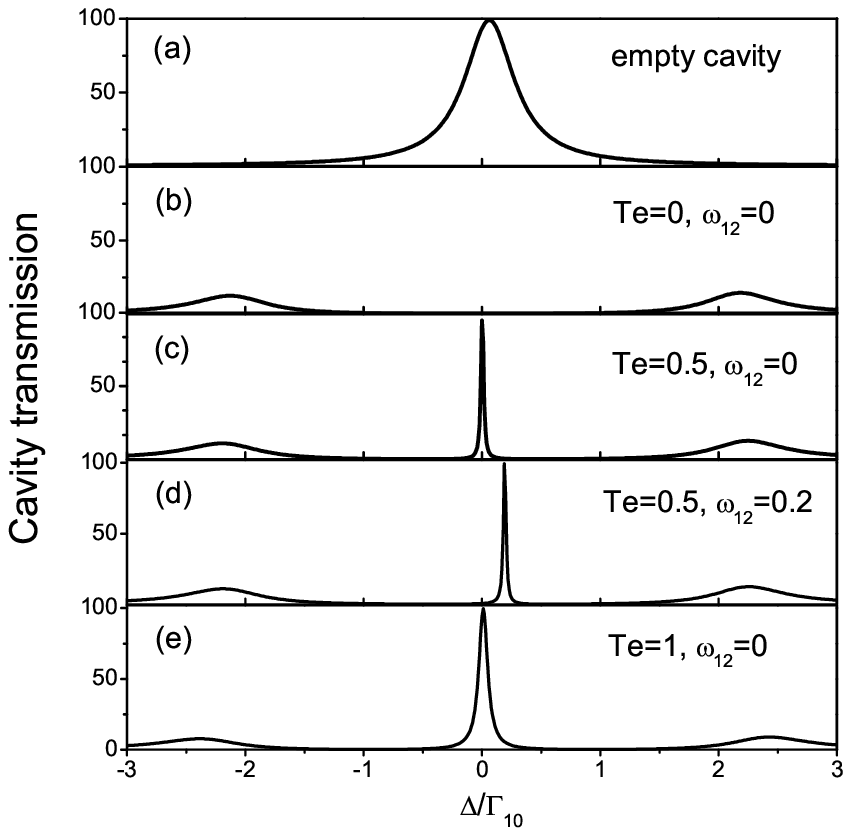}
\end{figure}

\clearpage Figure 3
\begin{figure}[htbp]
  \centering
  \includegraphics[width=9cm]{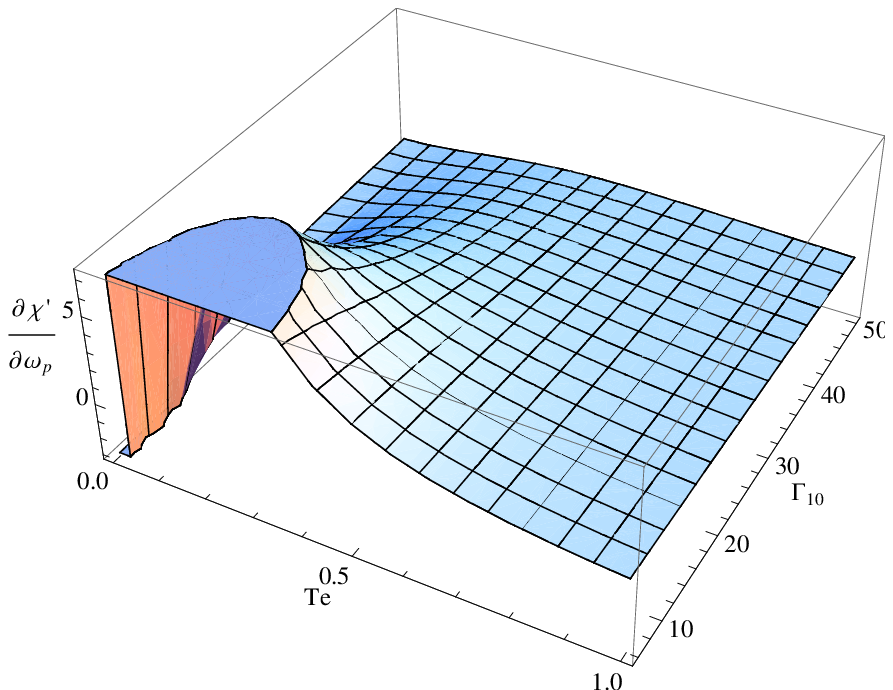}
\end{figure}

\end{document}